\documentclass[traditabstract]{aa}  

\usepackage{graphicx}
\usepackage{txfonts}
\usepackage{natbib}
\usepackage{epstopdf}
\begin{document}
   \title{A state-of-the-art analysis of the dwarf irregular galaxy NGC 6822  \thanks{Based on observations collected with the ACS on board the NASA/ESA HST;aaa}$^,$\thanks{Photometric data are available at the CDS via anonymous ftp cds.u-strasbg.fr (130.79.128.5)}}

\authorrunning{Fusco et al.}
\titlerunning{The SFH of dIrr galaxy NGC~6822}

   \author{F. Fusco     \inst{1,4}, R. Buonanno  \inst{1,2},  S. L. Hidalgo \inst{3,4}, A. Aparicio \inst{3,4}, A. Pietrinferni \inst{2},  G. Bono \inst{1,5},  M. Monelli \inst{3,4}, \and S. Cassisi \inst{2}
}

   \institute{Universit\`a di Roma Tor Vergata, Via della Ricerca Scientifica 1, 00133 Rome, Italy\\
              \email{federica.fusco@roma2.infn.it}
         \and
             INAF - Osservatorio Astronomico di Teramo, Via Mentore Maggini, 64100 Teramo, Italy  
         \and
            Instituto de Astrof\'isica de Canarias, Calle Via Lactea, E38205 La Laguna, Tenerife, Spain
         \and
           Departamento de Astrof\'isica, Universidad de La Laguna, Tenerife, Spain
           \and
             INAF - Osservatorio Astronomico di Roma, Via Frascati 33, 00040 Monte Porzio Catone, Italy  
              }

   \date{Received 2013; accepted 2014}

 \abstract{ We present a detailed photometric study of the
   dwarf irregular galaxy NGC 6822 aimed at investigating the
   properties of its stellar populations and, in particular, the
   presence of stellar radial gradients. Our goal is
   to analyse the stellar populations in six fields, which
   cover the whole bar of this dwarf galaxy. We derived the quantitative star formation history (SFH) of the six fields using the IAC method, involving
   IAC-pop/MinnIAC codes. The solutions we derived show an enhanced star formation rate (SFR) in Fields 1 and 3 during the past 500 Myr. The SFRs of the other fields are almost extinguished at very recent epochs and. We study the radial gradients of the SFR and consider the total mass converted into stars in two time intervals (between 0 and 0.5 Gyr ago and between 0.5 and 13.5 Gyr ago). We find that the scale lengths of the young and intermediate-to-old populations are perfectly compatible, with the exception of the young populations in Fields 1 and 3. The recent SF in these two fields is greater than in the other ones. This might be an indication that in these two fields we are sampling incipient spiral arms. Further evidence and new observations are required to prove this hypothesis. In addition, we derived the age-metallicity relations. As expected, the metallicity increases with time for all of the fields. We do not observe any radial gradient in the metallicity. }

\maketitle 

\defcitealias{fusco}{Paper I}	
\defcitealias{cannon}{C12}

\section{Introduction}
NGC 6822 is a barred dwarf irregular galaxy (dIrr) of type Ir IV-V
(\citealt{vandenbergh}) belonging to the Local Group (LG). This dwarf
galaxy has been the subject of many studies, since it is one of the
nearest dIrrs to the Milky Way (MW). 
The metallicity of NGC 6822 has been estimated by several authors
(e.g. \citealt{gallart, venn, sibbons}), with values ranging from
${\rm [Fe/H]}=-1.92\pm 0.35$ (\citealt{clementini}) to ${\rm
  [Fe/H]}=-0.49\pm 0.22$ (\citealt{venn}). \citet{kirby} carried out a spectroscopic study on single stars in seven dwarf irregular galaxies of the LG, for which they provide new and highly accurate metallicity estimates. The mean value obtained for NGC 6822 is ${\rm [Fe/H]}=-1.05$ with a dispersion of $0.49$~dex.

NGC 6822 is affected by a moderate foreground extinction because of
the low galactic latitude ($l=25.4^{\circ},~b=-18.4^{\circ}$,
\citealt{mateo}). According to Schlegel maps of the Galaxy
(\citealt{schlegel}, updated in \citealt{schlafly}), the reddening
along the line of sight towards NGC 6822 is ${\rm
  E(B-V)}=0.21$. Moreover, several authors have found a difference in
reddening between the innermost and outermost regions of NGC 6822
(\citealt{massey, gallart, gieren}). The estimates of the
distance cover a fairly wide interval ranging from ${\rm(m-M)_0=
23.31 \pm 0.02~ mag}$ (\citealt{feast}) to ${\rm(m-M)_0= 23.71\pm0.14~ mag}$
(\citealt{sc98}). In a recent study,
\citet[][hereinafter~\citetalias{fusco}]{fusco}, based on the
      {\textit{Hubble Space Telescope}} (HST) observations of
      \citet[][hereinafter~\citetalias{cannon}]{cannon}, estimated the reddening for the central region
       and external regions
      respectively as ${\rm E(B - V)_C = 0.37 \pm 0.02}$ and ${\rm E(B
        - V)_E = 0.30 \pm 0.03}$ and derived a distance modulus
      ${\rm(m-M)_0=23.54\pm 0.05~ mag}$, corresponding to $510\pm 10$~kpc.

The star formation history (SFH) of this galaxy was first studied by \citet{gall96b,gall96c}, who focused on the spatial variation of the stellar population in NGC 6822. Their analysis indicates that in the past 100--200 Myr an enhancement of the SF activity occurred over the whole main body of this galaxy. The strength of this enhancement has been different from one region to the other.
This was supported by \citetalias{cannon}, who studied the SFH of NGC 6822, based on deep HST data. They find that this galaxy formed more than $50\%$ of
the stars in the last ${\rm 5~Gyr}$. The SFH has been consistent throughout NGC 6822 in the past ${\rm
  500~Myr}$, with an increase in the star formation rate (SFR) in the last ${\rm 50~Myr}$.
  
NGC 6822 is well known for its complex
morphology. \citet{deblok00} mapped the entire extended distribution of H{~\scriptsize I} of NGC 6822, identifying a number of peculiar features. In
particular they observe an H{~\scriptsize I} cloud in the north-eastern region of this galaxy, and they suggest that this feature could be
associated with an interacting companion system. In addition, \citet{deblok03} noted
that the number of blue stars across the H{~\scriptsize I} cloud was particularly
large.

\citetalias{cannon} further analysed the nature of this H{~\scriptsize I} cloud. In their study the authors ruled out that the suspected putative companion, coincident with their Field 1, is a bona fide galaxy. This conclusion is based on the absence of an overdensity of red giant branch (RGB) stars in that region. However, the authors state that \textquotedblleft an interaction with a gas cloud without an ancient stellar component remains a viable solution\textquotedblright. 

The aim of the present work is to extend the analysis of \citetalias{fusco} in three more fields along the bar of NGC 6822, and thus end up with the total number of six fields. The photometric dataset is the same as for \citetalias{cannon}, but we perform a completely independent analysis, both in tools and method. In this paper the names of the fields coincide with those of \citetalias{cannon}. In particular, we investigate whether the nature of the suggested putative companion of NGC~6822 corresponding to Field 1 is confirmed. To our goal, we derive the SFH and age-metallicity relations (AMR) for the six fields, using the most updated version of the BaSTI evolutionary library\footnote{The full library is available at http://basti.oa-teramo.inaf.it.} (\citealt{basti}) and IAC-pop/MinnIAC population synthesis code (\citealt{IAC-pop, minniac}).

\begin{table}[t!]
\caption{List of the images in the three fields, with filter, exposure time (s), $\alpha$ (h), $\delta$ (deg).}             
\label{table:fields}      
\centering          
\begin{tabular}{c c c c c}     
\hline\hline       
                      
Field & Filter & Exp. time (s) & $\alpha$ (h) & $\delta$ (deg)\\
\hline                    
Field 1 & F475W &  423.0 &  19.726  &  $-$14.561\\
Field 1 & F475W &  463.0 &  19.726  &  $-$14.561\\
Field 1 & F814W &  883.0 &  19.726  &  $-$14.561\\
Field 1 & F814W &  463.0 &  19.726  &  $-$14.561\\
Field 2 & F475W &  559.5 &  19.732  &  $-$14.619\\
Field 2 & F475W &  559.5 &  19.732  &  $-$14.619\\
Field 2 & F814W &  559.0 &  19.732  &  $-$14.619\\
Field 2 & F814W &  559.0 &  19.732  &  $-$14.619\\
Field 3 & F475W &  559.5 &  19.738  &  $-$14.678\\
Field 3 & F475W &  559.5 &  19.738  &  $-$14.678\\
Field 3 & F814W &  559.0 &  19.738  &  $-$14.678\\
Field 3 & F814W &  559.0 &  19.738  &  $-$14.678\\

\hline                  
\end{tabular}
\end{table}

The paper is organized as follows. In Section~\ref{obs} the observations and dataset are described; in Section~\ref{stepop} we analyse the stellar populations identified in this galaxy; in Section~\ref{disc} we describe the SFH code, the input parameters, and the main results. The conclusions are presented in Section \ref{concl}.

\section{Observations and data reduction}
\label{obs}

   \begin{figure}
   \centering
   \includegraphics[scale=.48]{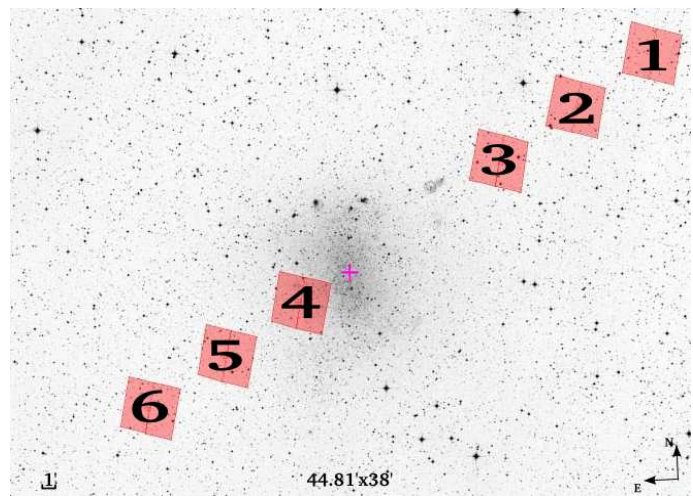}
   \caption{Position of the six fields with the associated label considered in the present paper. The cross indicates the position of the centre of NGC 6822, whereas the scale, the dimension of the field, and orientation are indicated. The position of Field 1 coincides with the H{~\scriptsize I} cloud identified by \citet{deblok00}.}
    \label{pointings}%
    \end{figure}

 \begin{figure}[t!]
   \centering
   \includegraphics[scale=0.5]{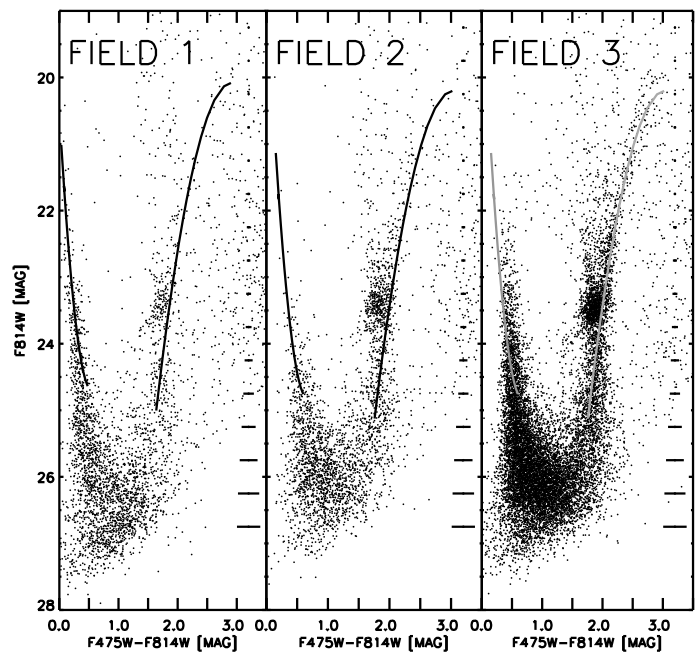}
   \caption{Final F814W, F475W$-$F814W CMDs, for Fields 1, 2, and 3.
     The associated photometric errors are also shown. Superimposed on the CMDs are
     the ridge lines of the CMD of Field 4 determined in
     \citetalias{fusco}.}
    \label{cmd_fin}%
   \end{figure}

The present photometry is based on archival HST datasets collected
with the Advanced Camera for Surveys (HST GO programme 12180,
P.I. Cannon). The same nomenclature as \citetalias{cannon} (Fig.~\ref{pointings}) is adopted here. In particular, we present here the analysis of three fields located in the north-western regions which are labelled Fields 1, 2, 3 from the outermost to the innermost. Field, filter, exposure time and coordinates
of our dataset are listed in Table \ref{table:fields}.

The images were prereduced using the standard HST pipeline. We
performed photometry on individual images using DAOPHOT
(\citealt{daophot}) and then simultaneous photometry on the four
images of each field using ALLFRAME (\citealt{allframe}). At the end
we obtained three catalogues -- one for each field, containing $\sim$15,000, $\sim$3,500, and $\sim$4,000 stars, respectively, from
the innermost to the outermost field. The data were transformed to the
standard VEGAMAG photometric system
(\citealt{sirianni}) using the most updated
zeropoints\footnote{Available at
  http://www.stsci.edu/hst/acs/analysis/zeropoints.}.

We thus obtained three F814W, F475W$-$F814W CMDs, for Fields 1, 2, and 3.
These CMDs are shown in Fig.~\ref{cmd_fin}.
The CMD of Field 4 (for a detailed analysis see \citetalias{fusco}) is given in the same figure.  We refer to \citetalias{fusco} for the CMDs of Fields 5 and 6.
  \begin{figure*}[t!]
   \centering
   \includegraphics[scale=0.7]{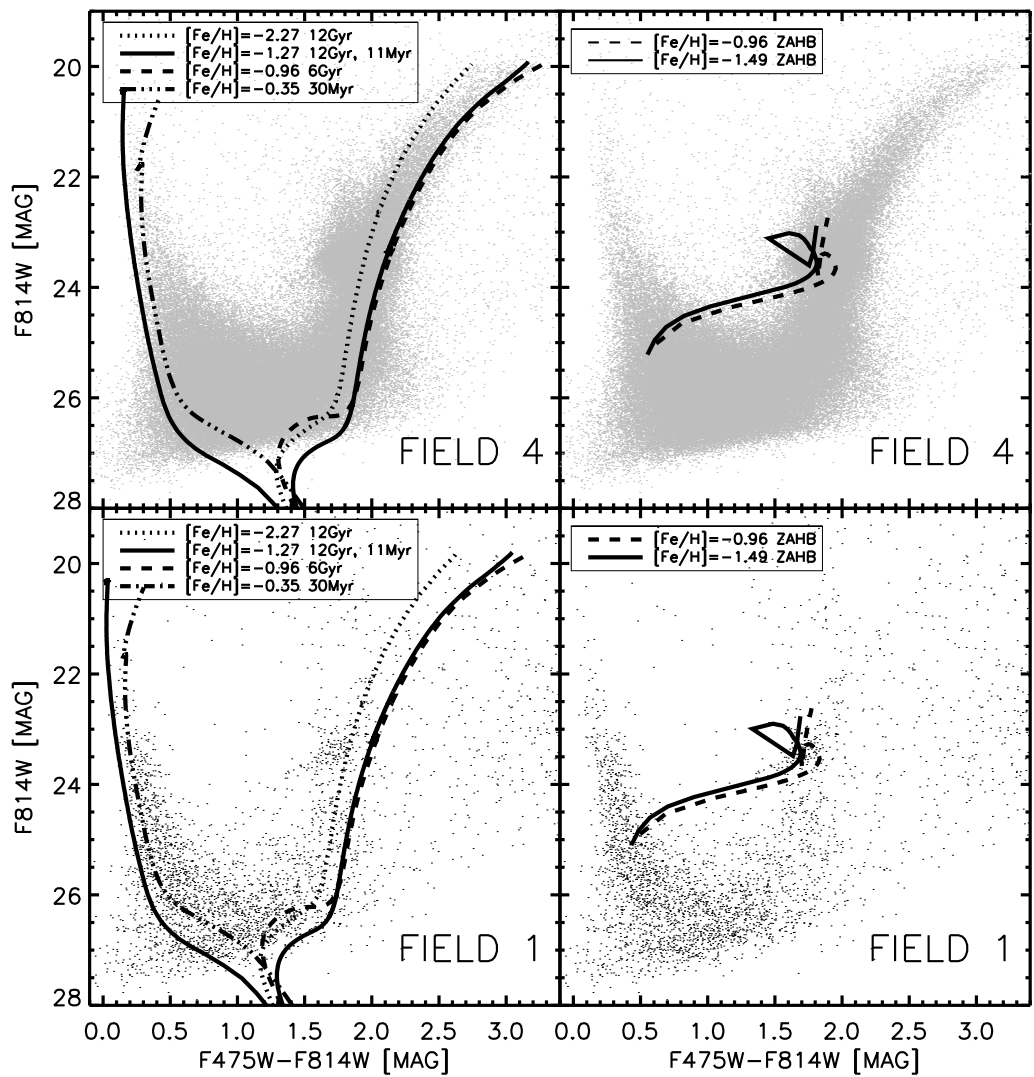}
   \caption{\textit{Left panel:} Comparison between the observed CMDs for Fields 4 (\textit{upper panel}) and 1 (\textit{lower panel}) and scaled solar theoretical isochrones for different assumptions about age and metallicity. Lines on the RGB are for ${\rm [Fe/H]=-1.27}$, 12 Gyr (continuous line), ${\rm [Fe/H]=-0.96}$, 6 Gyr (dashed line), and ${\rm [Fe/H]=-2.27}$, 12 Gyr (dotted line). Lines on the MS show two isochrones, respectively with ${\rm [Fe/H]=-1.27}$ and age 11 Myr (solid line) and with ${\rm [Fe/H]=-0.35}$ and age 30 Myr (dashed-dot-dot-dot line).  \textit{Right panel:} Comparison between core He-burning stars in the CMDs of Fields 1 and 4
and a theoretical ZAHB (with mass range between $0.6~{\rm M_{\odot}}$ and $2.8~{\rm M_{\odot}}$, in order to include both low-mass HB models and RC ones) for
two different metallicities, ${\rm [Fe/H]=-0.96}$, and ${\rm -1.49}$ (dashed and continuous line respectively). All the shown isochrones have been retrieved from the
BaSTI archive except the one corresponding to ${\rm [Fe/H]=-1.27}$ and age of 11 Myr, which has been computed
for the present work.}
    \label{iso}%
    \end{figure*}

The reddenings of Fields 4, 5, and 6 are given in \citetalias{fusco}.
To estimate the reddenings of the Fields 1, 2, and 3 we
superimposed on the CMDs the ridge lines of Field 4 already determined in
\citetalias{fusco} (solid lines in Fig.~\ref{cmd_fin}). Thus, the absolute reddening in Field 1 turns out to be the same as in Fields 5 and 6, whereas Fields 2 and 3 have the same absolute
reddening as Field 4. The last two lines of Table~\ref{table:ratio} show the distance modulus ${\rm(m-M)_{F814W}}$ and the E(F475W$-$F814W) reddening value for the six fields. We assumed the same reddening error of Fields 5 and 6 for Fields 1, 2, and 3.

\section{Preliminary findings on the stellar populations in NGC 6822}
\label{stepop}
To perform a preliminary analysis of the possible presence of different stellar populations across NGC 6822 we superimposed a set of isochrones to the CMDs of Fields 1 and 4 (i.e. the outermost and innermost fields). We adopted scaled solar isochrones from the BaSTI library (\citealt{basti}), and we used the distance modulus and reddening estimates discussed in Section~\ref{obs}. The result is shown in Fig.~\ref{iso}.

The comparison with the isochrones allows us to obtain some preliminary estimate of the age and metallicity of the stellar populations in NGC 6822:
\begin{itemize}
\item The two isochrones for ${\rm [Fe/H]=-1.27}$, age 11 Myr, and ${\rm [Fe/H]=-0.35}$, age 30 Myr, shown in Fig.~\ref{iso}, appear to constrain the blue edge of the main sequence (MS) fairly well. This comparison strongly suggests that NGC 6822 was still forming stars, at least in these two fields, in a very recent epoch.

\item The slope and the location of the RGB in the CMD, on the other hand, give a strong constraint on the metallicity. The RGB of NGC 6822 shows a significant spread, larger than the expected colour error at the magnitude of this branch, as shown in Fig.~\ref{cmd_fin}. We conclude that a spread in metallicity or a combined spread in age and metallicity is present in this dwarf galaxy. The slope of the RGB mainly depends on the metallicity, but we cannot exclude an age effect. Younger RGB stars attain bluer colours, thus mimicking a more metal-poor stellar population. From Fig. \ref{iso} (left panels) it appears that the blue edge of the RGB is represented well by the ${\rm [Fe/H]=-2.27}$ and 12 Gyr isochrone. On the other hand, the red edge is compatible with intermediate-age isochrones with slightly higher metallicity (i.e. ${\rm [Fe/H]=-1.27}$ and ${\rm [Fe/H]=-0.96}$). Therefore the age-metallicity degeneracy on the RGB does not allow us to constrain the spread in age and metallicity of the population in NGC 6822. The qualitative comparison between theoretical isochrones and observational data discloses that the photometric limit does not allow us to detect the MS turn off (TO) of the oldest population.

\item In the righthand panels of Fig.~\ref{iso} the zero age horizontal branch (ZAHB) is shown for two metallicities  ${\rm [Fe/H]=-0.96}$ and ${\rm [Fe/H]=-1.49}$ (mass range between $0.6~{\rm M_{\odot}}$ and $2.8~{\rm M_{\odot}}$).  The observed red clump appears well fitted by the central helium-burning sequence of ${\rm [Fe/H]=-0.96}$. There is some evidence that the intermediate-red colour HB is partially populated. The presence of the bright blue MS does not allow us to detect any blue HB stars.
\end{itemize}

\begin{table*}
\caption{Star counts and corresponding number ratios in the selected boxes of the CMD for the various
fields of NGC 6822. The distance modulus and the absolute reddening of the fields are listed.}              
\label{table:ratio}     
\centering                                      
\begin{tabular}{c c c c c c c}          
\hline\hline    
           &  Field 1   &   Field 2  &      Field 3  &   Field 4  &      Field 5   &   Field 6      \\
           \hline
Region      &  $ N_1 \pm \epsilon_{N1}$ &   $ N_2 \pm \epsilon_{N2}$ &   $ N_3 \pm \epsilon_{N3}$ &  $ N_4 \pm \epsilon_{N4}$ &   $ N_5 \pm \epsilon_{N5}$ &   $ N_6 \pm \epsilon_{N6}$   \\
       \hline
     LowMS &  $ 257 \pm 16   $ & $ 160    \pm  12  $ &  $1146  \pm   33 $ &  $  5047 \pm   71 $ & $  857   \pm  29  $ &  $ 477  \pm   21$ \\       
      UpMS &  $  86 \pm  9   $ & $  16    \pm   4  $ &  $ 188  \pm   13 $ &  $   779 \pm   27 $ & $   79   \pm   8  $ &  $  77  \pm    8$ \\
       LowRGB &  $ 106 \pm 10   $ & $ 219    \pm  14  $ &  $ 708  \pm   26 $ &  $ 10191 \pm  100 $ & $ 2435   \pm  49  $ &  $ 537  \pm   23$ \\
       UpRGB &  $  62 \pm  7   $ & $ 105    \pm  10  $ &  $ 403  \pm   20 $ &  $  6223 \pm   78 $ & $ 1313   \pm  36  $ &  $ 322  \pm   17$ \\
\hline
Ratio      &  $ R_1 \pm \epsilon_{R1}$ &   $ R_2 \pm \epsilon_{R2}$ &     $ R_3 \pm \epsilon_{R3}$&  $ R_4 \pm \epsilon_{R4}$&   $ R_5 \pm \epsilon_{R5}$&  $ R_6 \pm \epsilon_{R6}$ \\
\hline                               
    UpMS/UpRGB &  $ 1.4 \pm 0.2  $ & $  0.15   \pm 0.04  $ &  $ 0.47  \pm 0.03  $ &  $ 0.12  \pm 0.01 $ & $  0.06  \pm 0.01 $ &  $ 0.24  \pm 0.03  $ \\
   LowMS/LowRGB &  $ 2.4 \pm 0.2  $ & $  0.73   \pm 0.06  $ &  $ 1.56  \pm 0.05  $ &  $ 0.49  \pm 0.01  $ & $ 0.35  \pm 0.01 $ &  $ 0.89  \pm 0.04  $ \\
 LowMS/UpMS &   $  3.0\pm 0.5  $      &  $10.0\pm0.8$ &   $13.6\pm 0.4$     &      $  6.5\pm 0.1  $  & $10.8\pm0.4$ &  $6.2\pm 0.3$  \\
\hline
${\rm(m-M)_{F814W}}$   &  24.1$\pm$0.1 &24.2$\pm$0.1 &24.2$\pm$0.1 &24.2$\pm$0.1 &24.1$\pm$0.1 &24.1$\pm$0.1 \\                          
E(F475W$-$F814W)   &  0.51$\pm$0.05 & 0.63$\pm$0.05 &0.63$\pm$0.05 &0.63$\pm$0.04 &0.51$\pm$0.05 &0.51$\pm$0.05  \\                          
\hline     
\end{tabular}
\end{table*}

To have a closer insight into the nature of the stellar populations in the
different fields of NGC 6822 we performed star counts in the different
branches of the CMDs, including the three fields of
\citetalias{fusco}. We started by selecting five boxes in the most populated regions in the CMD of Field 4. These were defined as 
\begin{description}
\item[UpMS:] The upper MS box encloses MS stars
brighter than F814W${\rm  = 23.7~mag}$ where the MS luminosity function (LF) shows a
slight change in its slope.
\item[LowMS:] The lower main sequence box contains MS stars between ${\rm 25~mag <} $ F814W ${\rm  < 23.7~mag}$. According to the completeness tests we performed, F814W $ = 25$~mag is the limit of our photometry. The cut at F814W $ =23.7$~mag corresponds to the TO of the stars with an age of 450 Myr (for ${\rm [Fe/H]=-0.66}$). The comparison between UpMS and LowMS allows us to discriminate between stars younger and older than 450 Myr.
\item[RC: ] According to the change in slope of the RGB LF, we
 defined the interval ${\rm 23~mag\le F814W\le 23.9~mag}$ as the red
clump region. 
\item[LowRGB:] The lower RGB corresponds to ${\rm
  23.9~mag\le F814W\le 25~mag}$. The brighter limit in magnitude is set as the boundary of the RC region, while the fainter limit coincides with the limit in magnitude selected for our photometry according to the completeness tests.
\item[UpRGB:] Correspondingly we define the upper RGB branch as the region of the RGB brighter than the RC. The bluer part of the RGB was excluded with a by-eye inspection in order to minimize the contribution of the asymptotic giant branch to the star counts.

\end{description}

\begin{figure}
   \centering
   \includegraphics[scale=0.5]{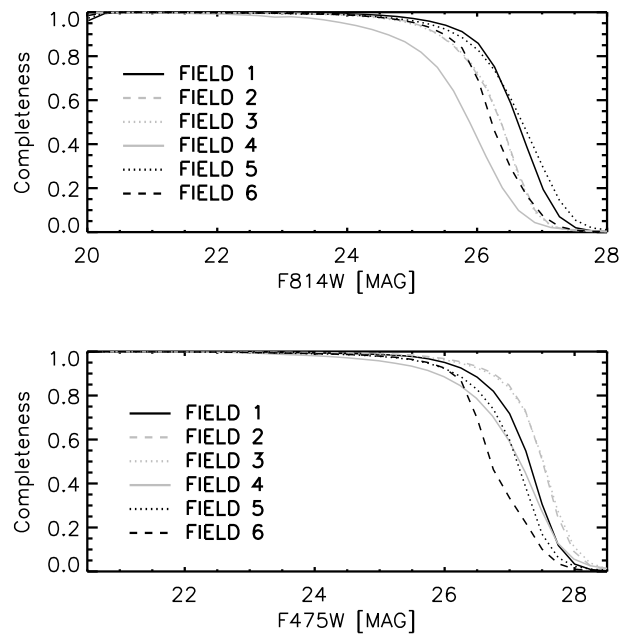}
   \caption{Result of the artificial star tests in all the fields for the two filters, F814W (\textit{upper panel}) and F475W (\textit{lower panel}). See text for a full description. }
    \label{crow}%
    \end{figure}

 Table~\ref{table:ratio} shows the star counts in
each box and the relative ratios. The errors listed in this table are given by the statistical fluctuations of the counts. We performed several experiments of
artificial stars in all the fields (for a complete description see Section~\ref{sec:sfh}) in order to confirm the statistical completeness of the star counts. The result is shown in Fig.~\ref{crow}, confirming that the counts at the
limit ${\rm F814W=25~mag}$ are complete at least at the
$85\%$ level.

To perform quantitative comparisons, the box boundaries were
shifted according to the relative reddening of each field. Inspection
of Table~\ref{table:ratio} immediately reveals that the two ratios
UpMS/UpRGB and LowMS/LowRGB, i.e. the number of stars on the MS with
respect to those on the RGB at the same luminosity level, turn out to
be definitely higher in Field 1, and marginally in Field 3, than in all the other fields. The completeness of the star counts in the boxes of the first ratios is nearly 100\%, hence, highly significative. These ratios, taken at their face
values, could provide a first hint that the stellar populations in
Field 1 and 3 could be peculiar owing to the underabundace of RGB stars with
respect to MS stars. In particular, this result could apparently suggest that a very recent burst of star formation took place in Field 1, which was more intense than in the other fields, including the innermost Field 4.

\begin{figure}
   \centering
   \includegraphics[scale=0.5]{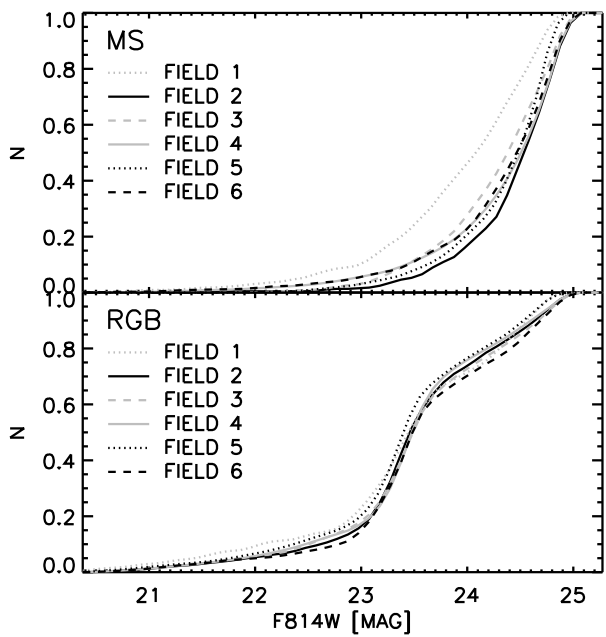}
   \caption{Cumulative LF of the MS (\textit{Upper panel}) and of the
     UpRGB (\textit{Lower panel}) for the six fields considered. The
     LF of the RGB in the interval ${\rm 22.7\le
       F814W\le23.7}$, corresponding to the RC region, has been obtained by interpolation along the RGB LF at magnitudes slightly fainter and brighter than the RC box.}
    \label{hist_tot}%
    \end{figure}

Therefore we computed the cumulative LFs of the RGB and of the MS for
the six fields, as shown in Fig.~\ref{hist_tot}. 
While the RGB LFs appear virtually identical in all of the
fields, the LF of Field 1 starts increasing at brighter magnitudes for the MS with
respect to all the other fields. This is a second hint to support
the idea that the star formation in Field 1 could somehow be peculiar
and that a relatively recent burst of star formation could have
occurred in this region.

\section{A detailed analysis of the SFH in the different fields}
\label{disc}

 The former findings indicate that significant differences may exist in the stellar populations of the six fields. To investigate these differences in depth, we use resolved population synthesis tools to calculate the SFH of the six fields. To obtain the SFH solution we considered only the stars with ${\rm \sigma_{F814W}}\le 0.2$, ${\rm\sigma_{F475W}}\le 0.2$, and ${\rm |sharpness| }\le 0.5$\footnote{The sharpness parameter is a standard DAOPHOT output, which quantifies the accuracy of the PSF profile.}. This selection allows us to obtain the solution using only stellar objects and thus improving the precision of the SFHs.

The first step is to accurately represent the observational effects which is done through tests of artificial stars (\citealt{tests}). We injected 10, 000, 000 stars in each image in 150 iterations. The stars were distributed uniformly in the images, and placed at a fixed relative distance set at $2~{\rm R_{PSF}+1pixel}$, where ${\rm R_{PSF}}$ is the PSF radius in pixels. Once the stars had been injected in the images, we performed a new photometry, with the same procedure as was used to reduce the original images. 

\subsection{Theoretical framework and input parameters}
\label{sec:sfh}

The SFH was derived by applying the
state-of-art package IAC-star/IAC-pop/MinnIAC
(\citealt{IAC-star,IAC-pop,minniac}), specifically developed for analysing resolved stellar populations. 
Even though the SFH of the various fields of NGC 6822
has been already studied by \citetalias{cannon}, the population synthesis
code we are using for the present analysis is completely independent
of the one adopted in the former analysis. At the same time, it is based
on an independent, updated library of stellar models.
Therefore, this investigation is also a test of the consistency of
SFH obtained with independent population synthesis codes and stellar
model libraries.

In short, IAC-star is used to create a synthetic CMD model. In our
case, the synthetic CMD model contains $10^7$ stars with a constant
star formation rate between 0 and 13.5 Gyr and uniform metallicity
distribution between $Z=0.0001$ and $Z=0.01$ -- corresponding to ${\rm [Fe/H]=-2.27}$ and ${\rm [Fe/H]=-0.27}$. The extreme values were selected to cover all the ages and all the metallicities (\citealt{clementini, venn}) of the stars present in NGC~6822. We selected the BaSTI
stellar evolution library (\citealt{basti}), a Kroupa initial mass function (\citealt{kroupa}), and a
30\% binary fraction. After a suitable simulation of the observational
effects in the synthetic CMD, the SFH is obtained using
IAC-pop/MinnIAC algorithms (\citealt{IAC-pop, minniac}) by comparison
of the distribution of stars in the synthetic CMD with that in the
observed CMD. The age bins selected are of 0.1 Gyr for ${\rm age<0.2~Gyr}$, 0.3 Gyr for ${\rm 0.2<age<0.5~Gyr}$, 0.25 Gyr for ${\rm 0.5<age<1~Gyr}$, 1 Gyr for ${\rm 2<age<5~Gyr}$, and 8.5 Gyr for ${\rm 5<age<13.5~Gyr}$. The metallicity bins selected are ${\rm 3\cdot10^{-4}}$ for ${\rm 10^{-4}\le Z\le10^{-3}}$, ${\rm 2\cdot10^{-3}}$ for ${\rm 10^{-3}\le Z\le3\cdot10^{-3}}$, ${\rm 4\cdot10^{-3}}$ for ${\rm 3\cdot10^{-3}\le Z\le7\cdot10^{-3}}$, and ${\rm 3\cdot10^{-3}}$ for ${\rm 7\cdot10^{-3}\le Z\le10^{-2}}$. These values were selected according to the age-metallicity resolution tests presented in Section~\ref{sfhamr}. With this method we obtain the smoothed SFHs, which include 24 single SFHs. For a full description we refer to \citet{minniac}.

To minimize the effect of the estimates in the external parameters, namely distance, reddening, and photometric zeropoints, IAC-pop/MinnIAC synthesis codes provide an offset both in magnitude and colour. These values are those which minimize the $\chi^2$ of the solution. Table~\ref{table:sol} shows the distance of the fields from the centre of NGC 6822, the $\chi^2$, and the offset in colour and magnitude of the solutions.

To evaluate the relative importance of the SFR in the fields, we evaluated the mass fraction of stars produced in two age bins. The results are listed in Table~\ref{table:sol}, where the old-intermediate- (between 13.5 and 0.5 Gyr ago) and young- (from 0.5 Gyr ago to date) epochs were taken into account. Inspection of this table reveals that the integral $\int_{0.5}^{13.5} \psi(t) dt$ in all the fields is comparable, whereas the integral $\int_{0}^{0.5} \psi(t) dt$ turns out to be very small for all the fields with the exception of Fields 1 and 3. In these fields the SFR shows a slight enhancement in this time interval. Nine and six times more stars were produced in Fields 1 and 3, respectively, than in Field 4.

In addition, the mean metallicity of the six fields is listed in Table~\ref{table:sol}. This result is perfectly compatible with the metallicity estimates by \citet{kirby}. Moreover, the associated spread in metallicity is compatible with the spectroscopic results.

\begin{table*}
\caption{Details of the SFH solution for the six fields. }
\label{table:sol}     
\centering                   
\begin{tabular}{c c cc cc   c c c}          
\hline\hline 
Field  &Dist. [deg]\tablefootmark{a}      &  $\chi^2$\tablefootmark{b}& ${\rm\Delta F814W}$\tablefootmark{c} &${\rm \Delta (F475W-F814W)}$  \tablefootmark{d}&  $\int_0^{0.5} \psi(t) dt $ \tablefootmark{e}&    $\int_{0.5}^{13.5} \psi(t) dt\tablefootmark{f} $& $\langle{\rm [Fe/H]}\rangle$ \tablefootmark{g} \\    
\hline
1     &  0.43    &   0.78  &$0.20  $&$0.05$  & $ 0.100 \pm0.007  $   &  $0.90\pm0.07 $  &  $-0.9\pm0.3$\\
2     &  0.32    &   0.76 &$0.20  $ &$0.15 $ & $ 0.017 \pm 0.002 $   &  $0.98\pm0.06 $  & $-0.8\pm0.1$ \\
3     &  0.22    &   1.04  &$0.10  $ &$0.05$ & $ 0.066 \pm 0.004 $   &  $0.93\pm0.08 $  &  $-0.9\pm0.2$\\
4     &  -0.06   &   2.35  &$0.08$&$0.15 $ & $ 0.011 \pm 0.001 $   &  $0.99\pm0.08 $  &  $-0.7\pm0.1$\\
5     &  -0.16   &   1.26  &$0.00  $&$0.07$ & $ 0.009 \pm 0.001 $   &  $0.99\pm0.07 $  &  $-1.0\pm0.2$ \\
6     &  -0.26   &   1.23  &$0.05$&$0.02$ & $ 0.018 \pm 0.002 $   &  $0.98\pm0.09 $  &  $-1.0\pm0.1$ \\

\hline                               
\end{tabular}
\tablefoot{
\tablefoottext{a}{Distance from the centre of each field, where the negative values identify the fields in the south-eastern region of the galaxy.}
\tablefoottext{b}{$\chi^2$ of the SFH solution.}
\tablefoottext{c}{Offsets in F814W of the solution.}
\tablefoottext{d}{Offsets in (F475W$-$F814W) of the solution.}
\tablefoottext{e}{Integral of the SFR between 0.5 and13.5 Gyr ago.}
\tablefoottext{f}{Integral of the SFR from 0.5 Gyr ago to date.}
\tablefoottext{g}{Mean metallicity and spread in metallicity of the solution.}
}
\end{table*}

\subsection{SFH solution and AMR}
\label{sfhamr}

 \begin{figure}
   \centering
   \includegraphics[scale=0.5]{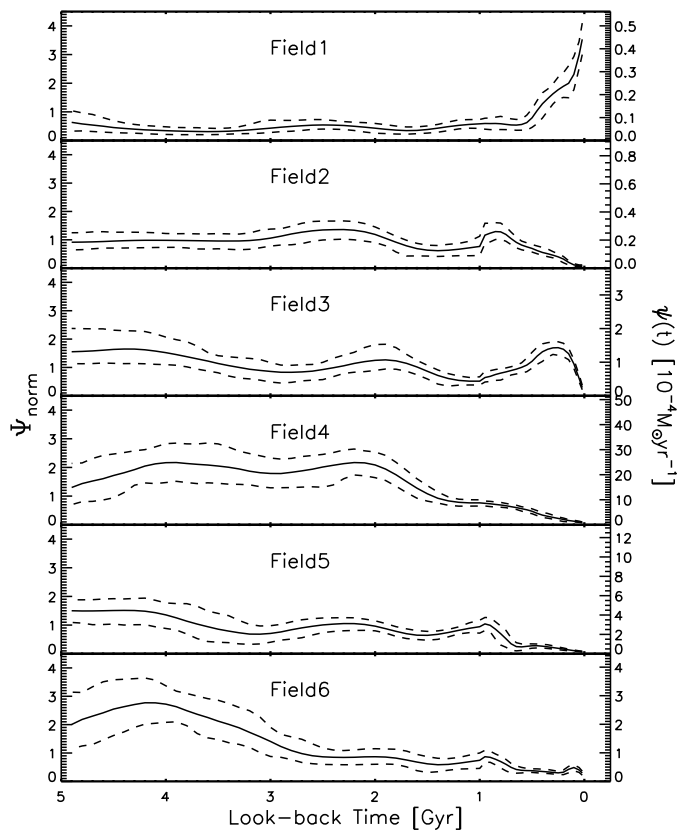}
   \caption{SFH for the six fields (solid line) as function of the time for the last 5 Gyr, with the relative error (dashed line). $\psi(t)$ (labeled of the right-side axis) is the stellar mass formed, while $\Psi_{\rm norm}$ (labeled of the left-side axis) shows $\psi(t)$ normalized to ${\rm \int_0^{13.5} \psi (t) dt}$. See text. }
    \label{sfh}%
    \end{figure}

   \begin{figure}
   \centering
   \includegraphics[scale=0.45]{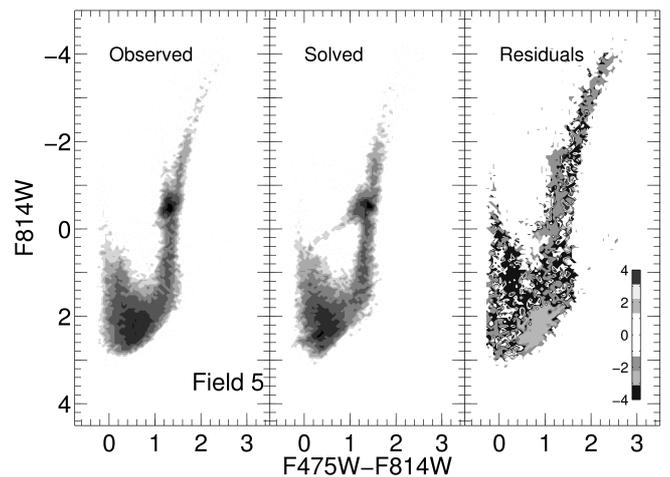}
   \caption{\textit{Left panel:} Observed CMD for Field 5. \textit{Central panel:} the CMD according to the SF produced by IAC-pop/MinnIAC. \textit{Right panel:} Residuals obtained as the difference between the observed and the solution CMDs counted in different bins in units of Poisson errors. Different colours correspond to different errors in agreement with the scale shown in the right panel. A similar result is obtained for all the other fields.}
    \label{residuals}%
    \end{figure}

   \begin{figure}
   \centering
   \includegraphics[scale=0.5]{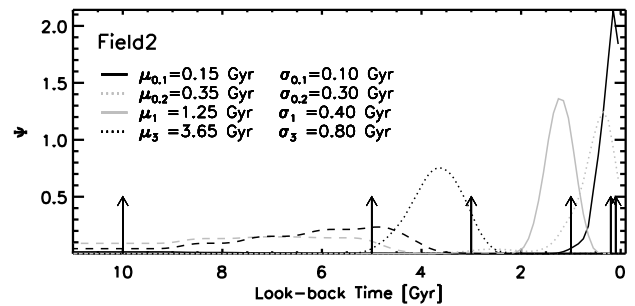}
   \caption{Age resolution tests for Field 2 for six different burst ages (indicated by the arrows), namely $10, 5, 3, 0.2, 0.1$ Gyr. Gaussian profiles represent the recovered SF episodes. The age peak and $\sigma$ are given. Each colour represents the recovered SF episode of a different age, in dashed-grey-, dashed-black-, dotted-black-, continuous-grey-, dotted-grey-, and continuous-black- lines respectively from the oldest to the youngest. Similar results are obtained for all the other fields.}
    \label{ageres}%
    \end{figure}

The SFH solution of the fields is shown in
Fig.~\ref{sfh}. The function ${\rm \psi(t)}$ is the stellar mass formed as a function of time, 
while $\Psi_{\rm norm}$ corresponds to ${\rm \psi(t)}$ normalized to ${\rm \int_{0}^{13.5} \psi (t) dt}$. The scales along the vertical axis are different for each field as a consequence of the difference in the total stellar mass. The solutions we obtained are consistent with the results presented by \citetalias{cannon}. Figure~\ref{residuals} shows observed, solution and residual CMDs in the case of Field 5.

The photometric limit avoids properly sampling the MS TO of the oldest population. To estimate the age limit of our solution, we carried out age resolution tests (for a full description we refer to \citealt{minniac}). Using IAC-star we created six mock populations of 10, 5, 3, 1, 0.2, 0.1 Gyr. Thus we performed the same procedure adopted to obtain the SFH solution, in order to recover the narrow mock bursts at different times. The result is shown in Fig.~\ref{ageres} for Field 2, as an example. Scrutiny of this figure shows that we are not able to recover the mock bursts of 5 and 10 Gyr. For this reason, the temporal scales of our solutions are not reliable for ages older than about 5 Gyr, although the integral ${\rm \int_{5}^{13.5} \psi (t) dt}$ is still accurate and well defined.

We now concentrate on the analysis of the individual SFH solutions (Fig.~\ref{sfh}) taking only ages $\le 5$~Gyr into account. This analysis discloses the following results:
\begin{description}
  \item[Field 1.] This field does not show any significant burst of SF until 500 Myr ago, when the most recent burst began and continued to increase up to a very recent epoch.
  \item[Field 2.] This field shows an SFR that remains quite constant over the whole time interval and tends to zero at the most recent epochs.
  \item[Field 3.] The SFR in Field 3 appears to be as constant as that of Field 2. In the past 500 Myr the SF, according to our solution, maintains a higher rate than in the other fields, but tends to disappear at very recent epochs.
  \item[Field 4.] A unique extended SF episode took place in this region of NGC 6822 between 5 and 1.5 Gyr ago. In the last 1.5 Gyr the SFR vanished.
  \item[Field 5.] The SF calculated for this field is analogous to the rate obtained in Field 2. 
   \item[Field 6.] The SF calculated for this field is analogous to the one obtained in Field 4.
\end{description}

From the analysis of the single solutions of the SF, we find that in Fields 2, 4, 5, and 6, the SFR has been slowly decreasing in the last 1.5 Gyr, and it is exhausting at very recent epochs. Fields 1 and 3 are apparent exceptions to this frame. For these field the solutions show an enhancement in the SFR in the last 500 Myr, and in the case of Field 1 it has still been active at very recent epochs.

\begin{figure}
   \centering
   \includegraphics[scale=0.5]{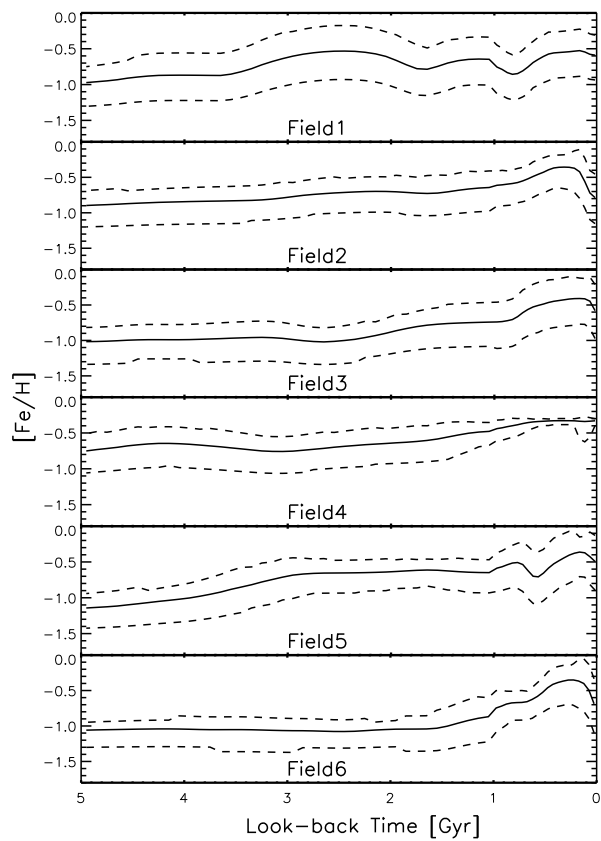}
   \caption{AMRs for the six fields (solid lines), including error bars (dashed lines), as function of the time for the last 5 Gyr.  }
    \label{zmedage}%
    \end{figure}

In addition, the IAC-pop/MinnIAC method allows deriving the AMR as a part of the SFH. The AMRs for six fields of NGC 6822 are shown in Fig.~\ref{zmedage} for the last 5 Gyr.

The AMRs disclose that the metallicity grows with time in all the six fields, ranging between ${\rm -1.0 \lesssim [Fe/H] \lesssim -0.5}$. According to our solution no difference in the AMR of the six fields is identified. These values are compatible with the mean metallicity and the spread in metallicity recently found spectroscopically for NGC 6822 by \citet{kirby}.

   \begin{figure}
   \centering
   \includegraphics[scale=0.5]{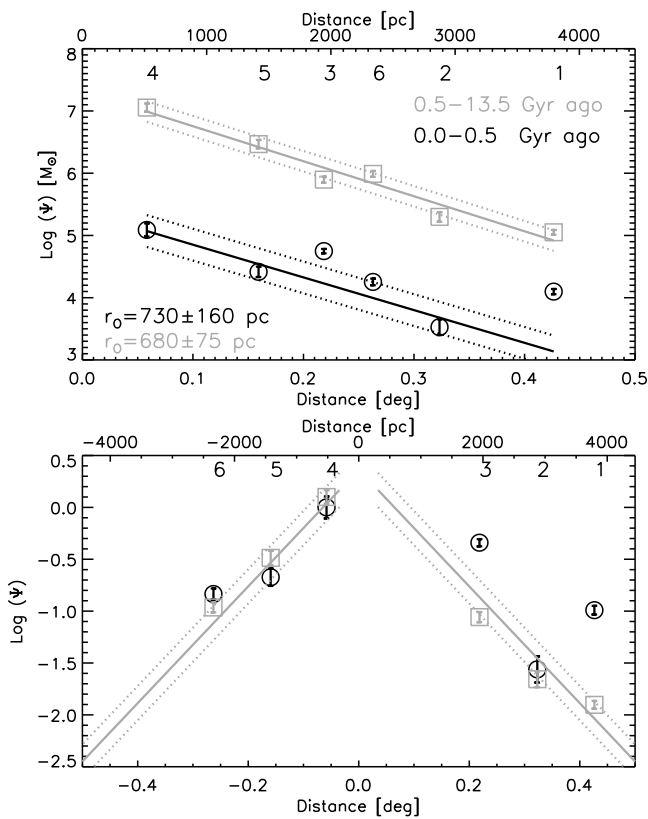}
   \caption{The integral of the SFR, ${\rm \psi(t)}$, for the young (from $0.5$ Gyr to date; black dots) and intermediate-to-old (between 0.5 and 13.5 Gyr ago; grey squares) stellar populations, as a function of the galactocentric distance. \textit{Upper panel}: the absolute value of the galactocentric distance is shown in the horizontal axis. Distances are given both in degrees (lower axis scale) and in parsecs (upper axis scale). Distances in parsecs have been obtained assuming a distance to NGC 6822 of 510 kpc (\citetalias{fusco}). The solid lines show the best fit of the points in the two time intervals, with the associated errors (dotted lines). Fields 1 and 3 were excluded from the fit in the case of the young stellar population. The resulting scale length ${\rm r_0}$ and the name of the field are indicated. \textit{Lower panel}: same as the \textit{upper panel}, but the galactocentric distances are shown as they appear along the galactic semi-major axis (negative and positive values corresponding to eastern and western fields, respectively). Points have been shifted vertically in order to make the older and younger distribution barycentres coincide. The vertical scale zeropoint is arbitrary. Grey lines reproduce the fit shown in the \textit{upper panel} for the older population. The colour coding is the same a in the \textit{upper panel}.}
    \label{zmeddistance1}%
    \end{figure}

   \begin{figure}
   \centering
   \includegraphics[scale=0.5]{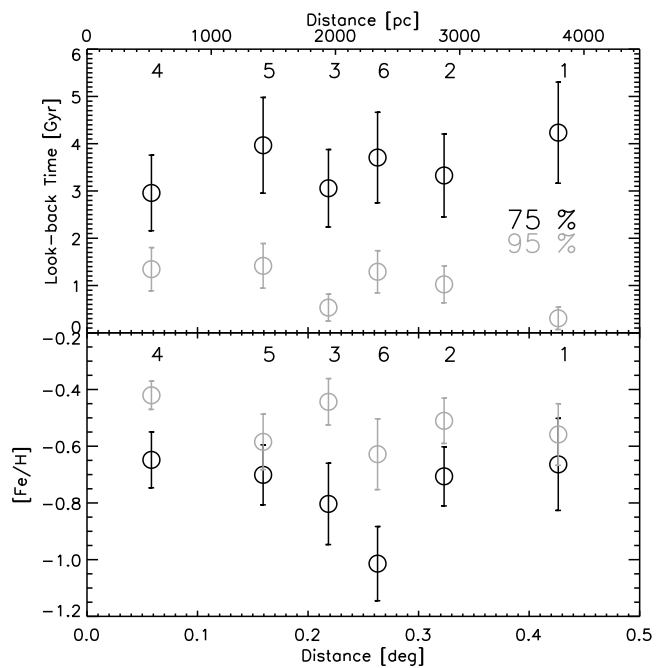}
   \caption{Age and metallicity as a function of the galactocentric distance. The latter is shown both in degrees (lower axis scale) and in parsecs (upper axis scale). The distances in parsecs have been obtained assuming a distance to NGC 6822 of 510 kpc (\citetalias{fusco}). \textit{Upper panel}:  look-back times of the 75 and 95 percentiles of the SFR  as a function of the galactocentric distance. Numbers from 1 to 6 within the panel identify the corresponding fields. \textit{Lower panel}: the mean metallicity for the same percentiles. The percentiles and the colour coding are the same as in the \textit{upper panel}. }
    \label{zmeddistance}%
    \end{figure}

\subsection{Radial gradients}

Our dataset samples the bar of NGC 6822 up to about five times its scale length. This offers a good opportunity to investigate the radial stellar populations distribution and the possible presence of a population gradients.

Several authors have studied the population gradients in the dwarf
galaxies of the LG (e.g. \citealt{hidalgo13} and references
therein). In all the cases the trend we found is the same: the age of the youngest populations gets older for increasing radius, and progressively
younger populations are only found in the innermost regions.  The
observed effects agree with models of simulated dwarf galaxies (\citealt{schroyen}). 

Regarding NGC 6822, the spatial distribution of the stellar populations have already been discussed by \citet{gall96b,gall96c}. They studied the
central regions of the galaxy and found an enhancement of the SFR
in the past 100-200 Myr, which occurred over the whole main body of the
galaxy. The strength of this enhancement is different between the
different regions of NGC 6822. This results were confirmed by \citetalias{cannon}, who noticed that NGC 6822 formed more than 50\% of the stars in the last 5 Gyr.

In this research we started evaluating ${\rm\int \psi (t)~dt}$ in two time intervals, between 0 and 0.5 Gyr ago and between 0.5 and 13.5 Gyr ago, as a function of the galactocentric distance. The results are shown in Fig.~\ref{zmeddistance1}. Clear exponential profiles are disclosed in all cases, except for the young populations of Fields 1 and 3. The scale lengths are compatible within the errors for the old and young populations. These result in $ r_{0.5}^{13.5}=680\pm75~{\rm pc}$ and $ r_0^{0.5}=730\pm160~{\rm pc}$ for the old- and the young-population. $r_0^{0.5}$ was computed excluding Fields 1 and 3. Fields 1 and 3 show young populations that exceed the ones of the surrounding fields by a factors of 10 and 3, respectively. This may be the simple result of a stochastic enhancement of the SFR in recent epochs. However, this could also be the signature of incipient spiral arms. With the present dataset this is only a speculative suggestion. New data are required to either support or rule out this possibility.

Following \citet{hidalgo09}, we consider the times corresponding to the 75 and 95 percentiles of the SFR to further analyse the radial gradients. In other words, the times $t_{75}$ such that ${\rm \int_0^{t_{75}}\psi(t)~ dt/\int_0^T\psi(t)\ d t=0.75}$ and $t_{95}$ such that ${\rm \int_0^{t_{95}}\psi(t)\ d t/\int_0^T\psi(t)\ d t=0.95}$, where $T$ is the current age of the system. Figure~\ref{zmeddistance} shows the look-back time and the metallicity as a function of the galactocentric distance for the two percentiles. Scrutiny of this figure reveals that the 75 percentile corresponds to similar look-back times in all the cases. The look-back times associated to the 95 percentile are similar for Fields 2, 4, 5, and 6, but they correspond to a more recent epoch for Fields 1 and 3. Moreover, Fig.~\ref{zmeddistance} discloses that the metallicity increases in all the fields from the 75 to the 95 percentiles, but it does not show any radial gradient. To quantify these findings, we obtained an empirical relation $Z=Z(t,D)$, where $Z$ is the metallicity, \textit{t} is the look-back time, and \textit{D} the galactocentric distance in kpc

\begin{equation}
Z=[(5.7\pm 1.5)-(0.57\pm 0.08)t-(0.8\pm 8.5)D]\cdot10^{-3}.
\end{equation}

The errors of the fitting parameters confirm that the metallicity significantly increases with time, but that no significant metallicity gradient as a function of the distance is present. These results confirm the findings that have already been presented.

NGC 6822 does not show the trend found in similar galaxies by \citet{hidalgo13}. They studied the SFH of two dIrr/dSph galaxies, namely LGS-3 and Phoenix, whose total stellar masses are about 100 times lower than that of NGC 6822. They find that, for ages $<9$ Gyr, the SFR gradually decreases outwards and that the metallicity does not change significantly with time or with distance from the centre. For recent epochs they find a steep increase in the metallicity, which does not have a counterpart in the SFR.

\section{Conclusions}
\label{concl}
In this paper we have presented a new photometric analysis of three fields in the north-eastern region of the dIrr NGC 6822. We carried out a new and independent study of the SFH throughout this galaxy, considering both these three fields and the three fields presented in \citetalias{fusco}. The study was based on the IAC method, involving the IAC-pop/MinnIAC codes. As a consequence this work presents several new results for the SFRs which deepen some of  the previous findings by \citetalias{cannon}. In addition, as an important new result, we provide the AMRs as a function of time.

The performed analysis provided the following results.

\begin{itemize}

\item The SFH obtained using this methodology reveals that Field 1 shows an enhancement in the SFR that began around 500 Myr ago and has been increasing up to a very recent epoch. The SFR during this time interval is twice as high as the average for that field at earlier epochs. In all the other fields we find that the last Gyr was characterized by a slow decrease in the SFR. The AMRs show that the metallicity grows with time throughout the whole galaxy. The mean metallicity and the metallicity spread supports recent spectroscopic measurements by \citet{kirby}.
  
\item We studied the radial gradients of the SFR. We considered the total mass converted into stars for young (between 0 and 0.5 Gyr ago) and intermediate-to-old (between 0.5 and 13.5 Gyr ago) ages. We observe a clear decreasing trend with the galactocentric distance in both cases, with very similar scale lengths. Fields 1 and 3 are an exception. The recent SF in these two fields clearly exceeds the one of the other fields. Nevertheless the intermediate-to-old populations follow the same trend as in the other fields. This can be considered as a sign that the intermediate-old stellar population is mixed well along the bar of NGC 6822, while this does not hold for young populations. Thus, young populations follows the same trend with the same scalelength as the intermediate-to-old ones in all but two fields (Fields 1 and 3). This may indicate that in these two fields we are sampling incipient spiral arms. Nevertheless, this conclusion has to be considered as only hypothesis as long as we do not have further evidence from deeper observations. 

\item The metallicity does not show any gradient along the bar, regardless of the SFR gradients and fluctuations described above. However, the metallicity increases significantly with time all along the bar. 
\end{itemize}

\begin{acknowledgements}
We warmly thank the anonymous referee, who helped to improve the presentation of this paper. FF thanks the IAC for scientific and financial support for the present work, and Giacinto Iannicola and Ivan Ferraro (INAF OAR) for the help in computing the metallicity-age-distance relation. We acknowledge financial support from the Italian National Institute of Astrophysics and the Ministry of University and Research (grants PRIN-INAF 2011 and PRIN-MIUR 2010LY5N2T), from the IAC (grant 310394) and the Spanish Ministries of Science and Innovation and of Economy and Competitiveness (grants AYA2007-3E3507 and AYA2010-16717).

\end{acknowledgements}

\end{document}